\documentclass[aip,graphicx,amsmath,amssymb,reprint]{revtex4-1}

\usepackage{graphicx}
\usepackage{dcolumn}
\usepackage{bm}

\usepackage[utf8]{inputenc}
\usepackage[T1]{fontenc}
\usepackage{mathptmx}
\usepackage{etoolbox}
\makeatletter
\def\@email#1#2{%
 \endgroup
 \patchcmd{\titleblock@produce}
  {\frontmatter@RRAPformat}
  {\frontmatter@RRAPformat{\produce@RRAP{*#1\href{mailto:#2}{#2}}}\frontmatter@RRAPformat}
  {}{}
}%
\makeatother

\begin{document}


\title{Compact computer controlled biaxial tensile device for low-temperature transport measurements of layered materials} 

\author{S.V. Zaitsev-Zotov}
\email[]{serzz@cplire.ru}
\affiliation{Kotelnikov Institute of Radioengineering and Electronics of RAS, Mokhovaya 11, bld. 7, Moscow 125009, Russia}

\affiliation{Physics Department, HSE University, 20 Myasnitskaya ulitsa , Moscow 101000, Russia}


\begin{abstract}
A biaxial tensile device for the transport study of layered materials is described. The device is mounted on the standard 24 pin zero force connector and can be moved between various setups. The compact design of the device makes it suitable for a wide range of studies. In our case, it is placed inside a 50 mm diameter chamber in the cryocooler and is used in the temperature range 9–310 K. A sample is glued in the
center of a polyimide cruciform substrate, the ends of which are connected to a tension system driven by four computer-controlled stepper motors providing tensile force up to 30 N. Computer simulation results and their experimental verification show that tensile strain along one
axis depends on the tensile load along the perpendicular direction, and this dependence turns out to be relatively strong and exceeds 40\%. The operation of the device is demonstrated by studying the effect of deformation on the electrical conductivity of the layered compound 2H-NbS2.\\
\\
https://doi.org/10.1063/5.0187818
\end{abstract}


\maketitle 

\section{Introduction}
Straintronics is a rapidly growing field of research \cite{Roldan_2015,frisenda2017biaxial,Peng_2020}. By applying uniaxial strain, the properties of various materials can be varied from metallic to dielectric, and the various types transitions (superconducting, Peierls, magnetic {\it  etc.}) can be suppressed or initialed \cite{Tritt,Chi_2023}. The biaxial strain offers even more possibilities if it provides the possibility of manipulating the crystalline lattice parameters in two directions. For instance, by applying strain along one or another direction, the direction of the wave vector of the charge-density waves in TbTe$_3$ can be turned by $90^\circ$ \cite{Gallofrantz2023}. Biaxial strain devices are known in mechanics of constructive materials \cite{makinde1992development}, whereas a very limited number of biaxial stain/stress devices suitable for nanoscience applications are described in the literature. Geandier and coauthors \cite{Geandier} described a biaxial tensile machine working in the synchrotron environment in which an elastic cruciform substrate (polyimide) with deposited film under study is mechanically stretched by four motors connected to the cross ends and controlling bidirectional strain components. Pasquier and coauthors \cite{Pasquier_2023} described a tunable biaxial strain device in which the strain is created by pressing a table with four pushers close to the center of an elastic cross with a sample. The strain produced by this setup is equal for both directions. More complex version of bidirectional strain device analogous to one of Ref.~\cite{Geandier} is described by Gallo-Frantz and coauthors \cite{Gallofrantz2023}. Here the sample is glued in the center of a polyimide cross pressed by tensile strain to a cold finger of a flow cryostat inside a vacuum chamber. The strain of polyimide cross is provided by four motors, the straining force being measured by four force sensors. This device allows to control the stress along two directions of the cross independently. Various modifications of this device were adopted for different types of measurements including X-ray study, and transport measurements. Here we described a compact device which has strain controls along two transverse directions and can be transferred between different experimental environments keeping sample deformation. In our case it is adapted into cryocooler for transport measurements in the temperature range 10 - 310 K.

\section{Compact Biaxial Tensile Cell Design}
The compact biaxial tensile strain device 
is somewhat similar to 
ones described in earlier publications \cite{Geandier,Gallofrantz2023,Pasquier_2023}, but has different technical realization. In contrast to devices described in mentioned papers, we control the substrate strain rather than the stress applied.

\begin{figure*}
\includegraphics[width=0.3\textwidth]{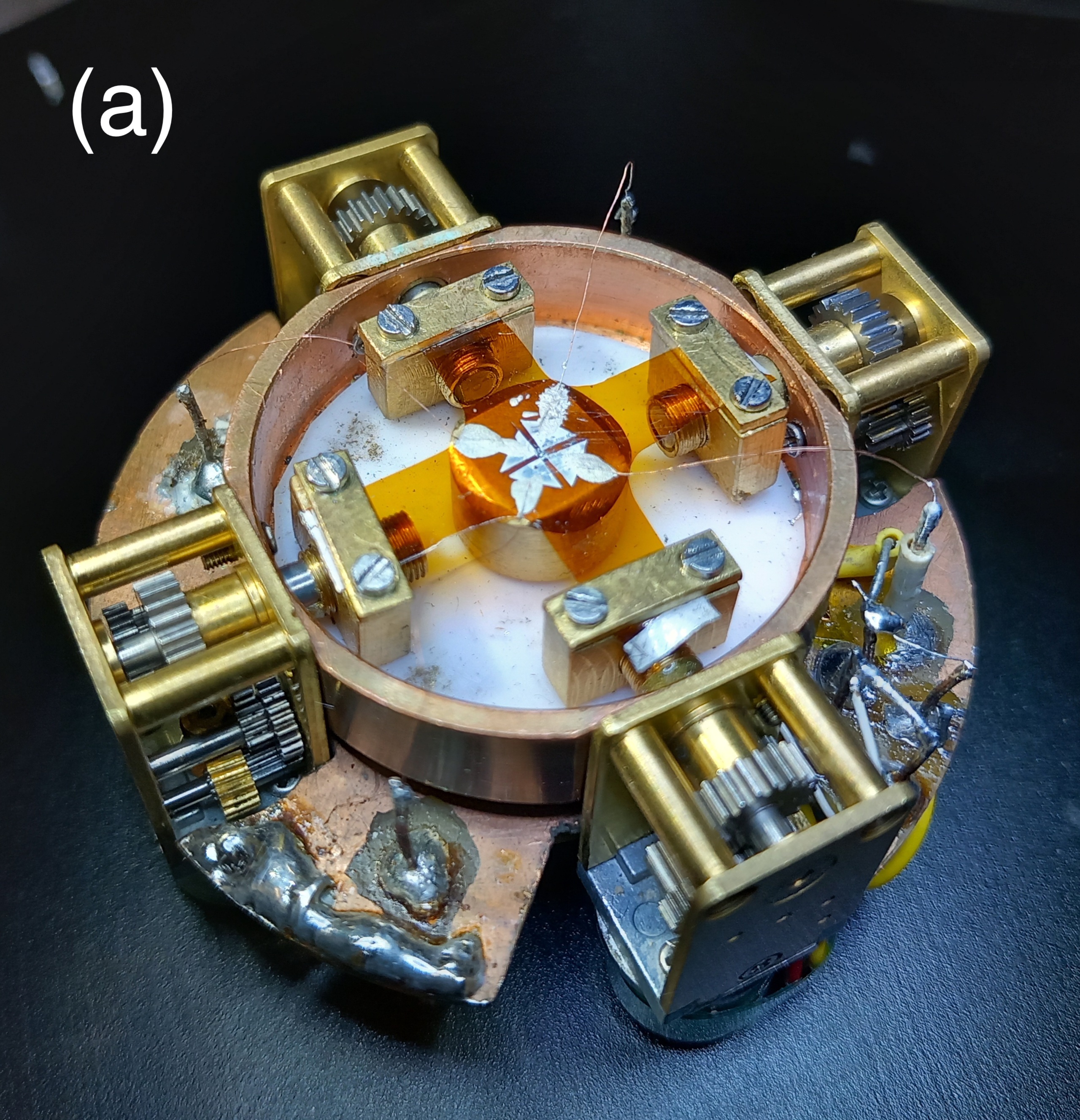} \ \ 
\includegraphics[width=0.3\textwidth]{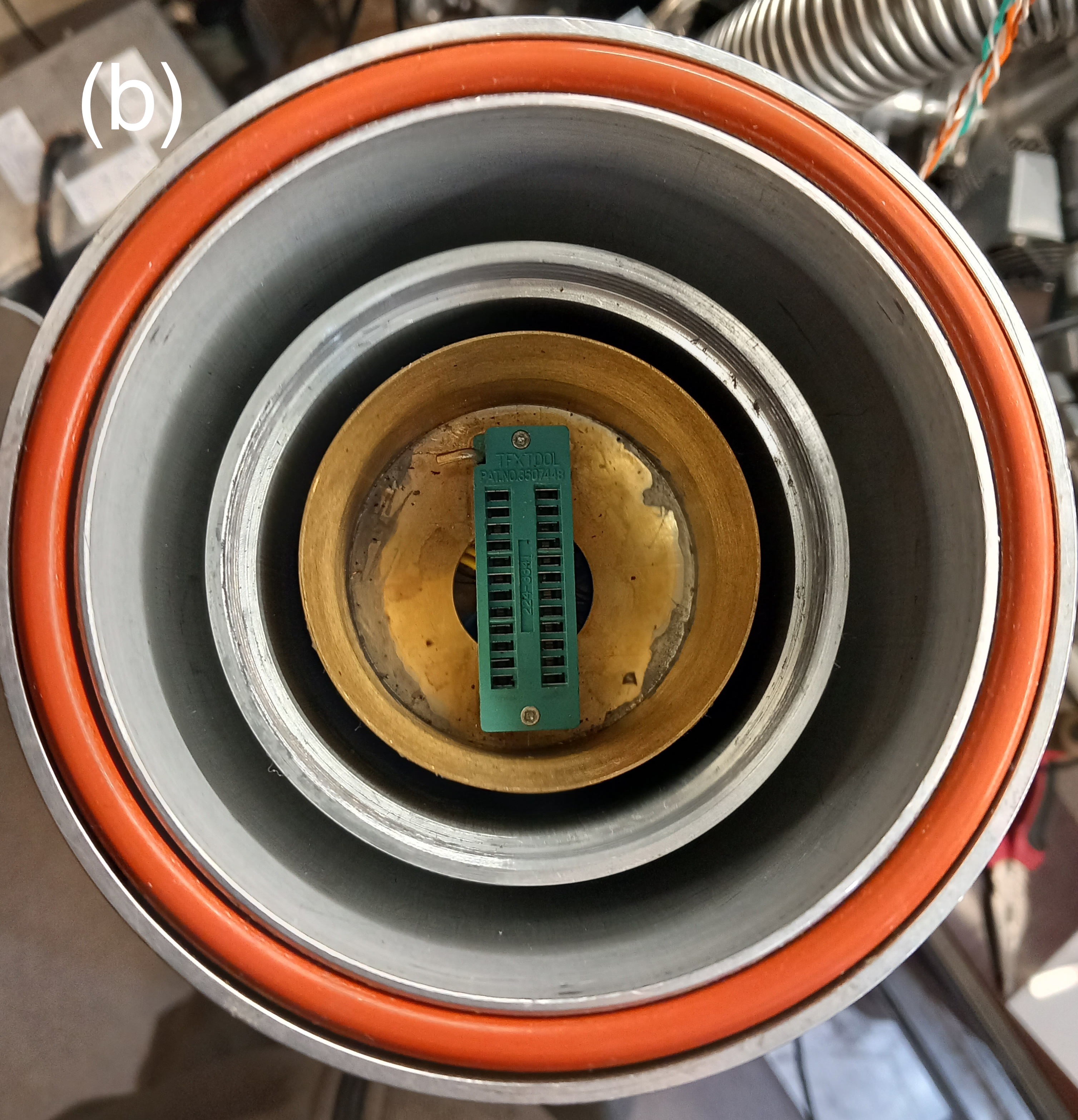} \ \ 
\includegraphics[width=0.3\textwidth]{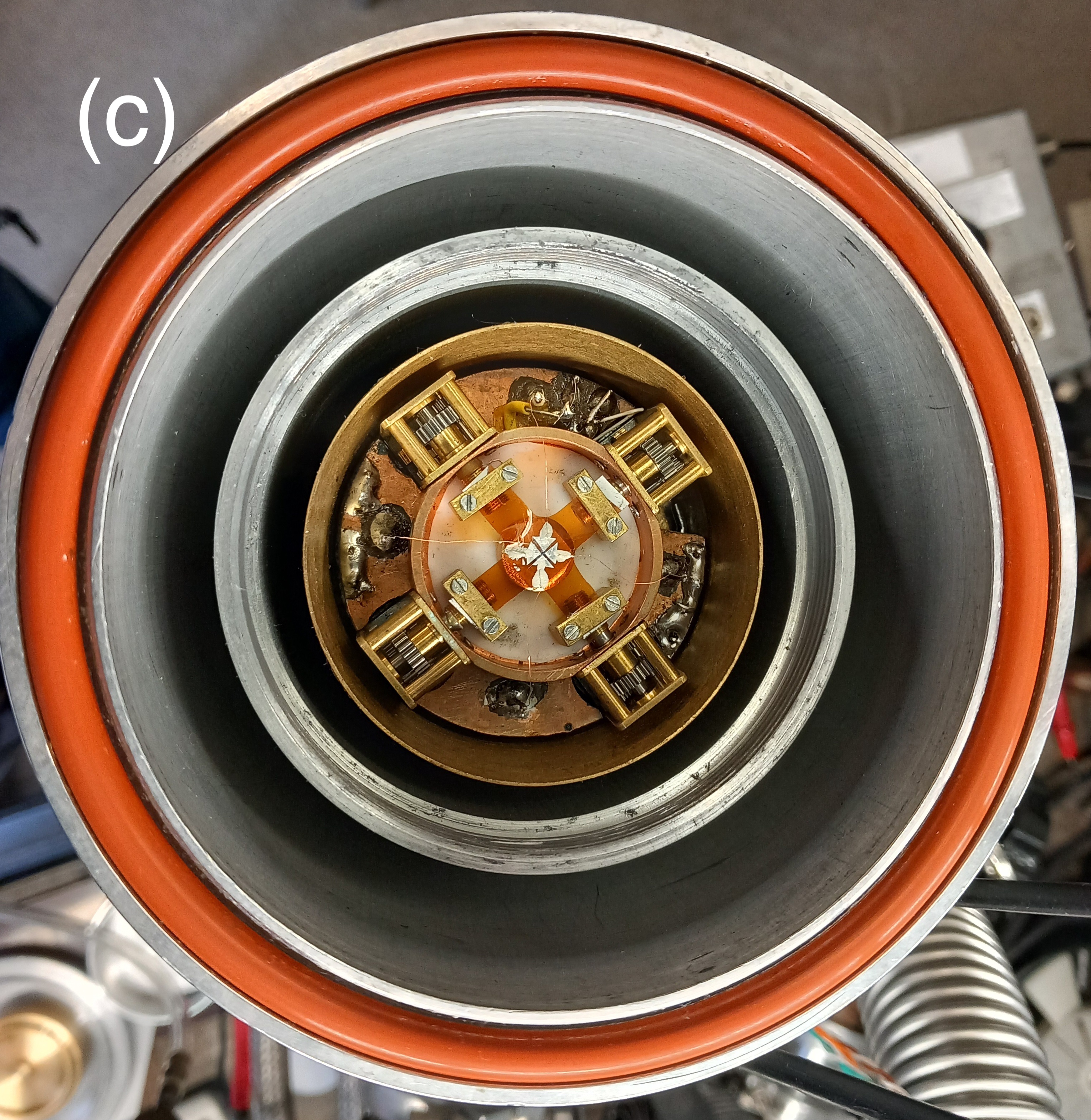}
\caption{a) Tensile cell with a sample mounted; b) Zero-force connector inside a cryosystem for cell installation; c) Cryosystem with the tensile cell installed. The caps of the radiation screen, cryostat and cell chamber are removed.
\label{fig:both}}%
\end{figure*}

\begin{figure}
\includegraphics[width=0.5\textwidth]{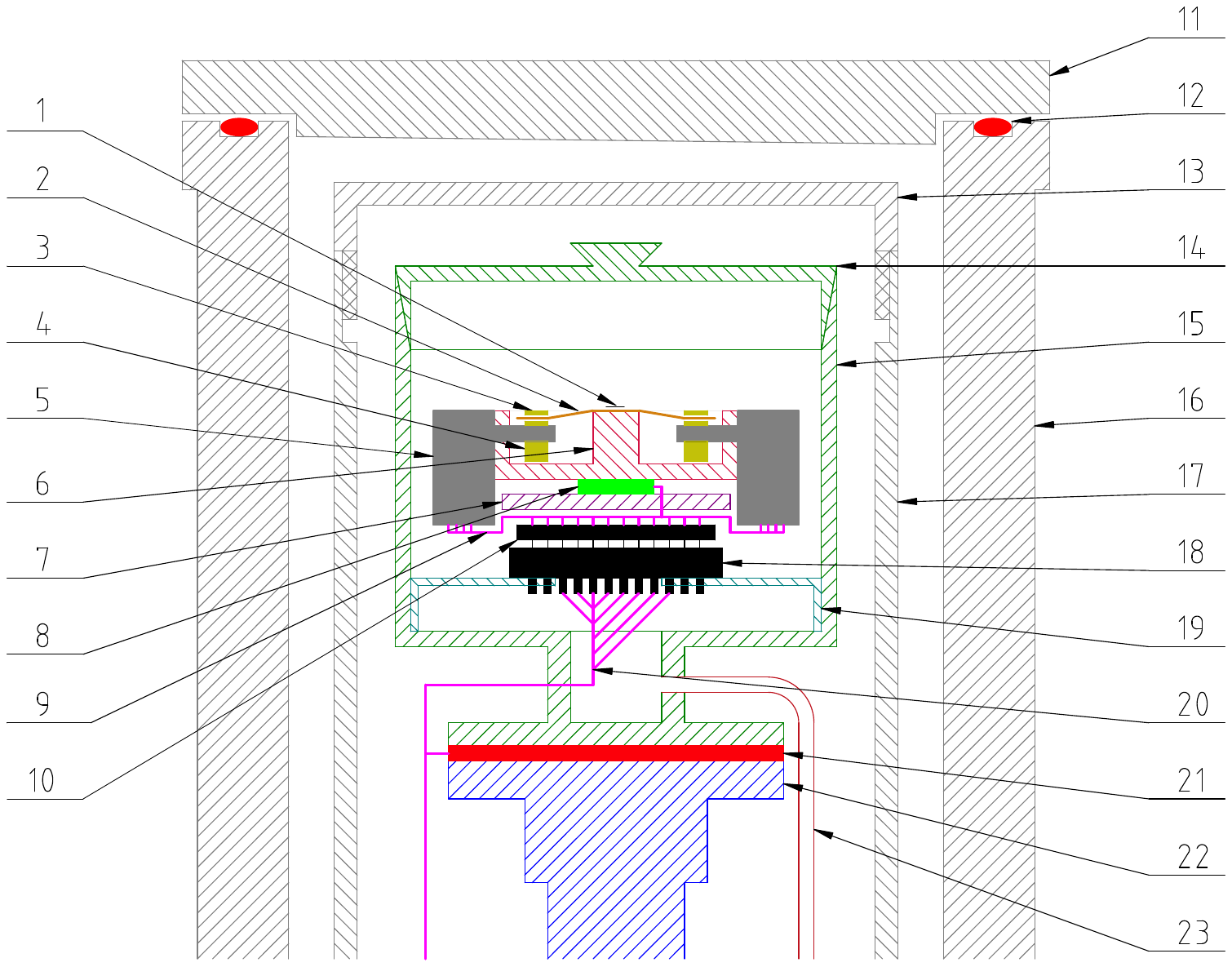} 
\caption{Schematic diagram of the cryogenic setup with 2D tensile cell installed. The numbers mark: 1 - sample; 2 -  polyimide substrate; 3 - brass clamp; 4 - moving brass block with thread; 5 - stepper motor with gearbox; 6 - copper pot; 7 - foiled fiberglass disk; 8 - temperature sensor; 9 - 2D tensile cell wiring; 10 - 24 pin panel; 11 - vacuum cup; 12 - Viton O-ring; 13 - radiation screen cup; 14 - chamber lid; 15 - chamber; 16 - cryostat wall; 17 - radiation screen wall; 18 - zero force 24 pin connector; 19 - internal connector mount; 20 - wiring; 21 - heater; 22 - second stage of cryocooler; 23 - heat exchange gas/pumping tube.
\label{fig:cryo}}%
\end{figure}

Figure~\ref{fig:both} shows photos of the tensile cell (a) and top views of the cryogenic system without (b) and with the tensile cell installed (c), and Figure~\ref{fig:cryo} shows the schematic diagram of the system. All the numbers in the further description correspond to components shown in Figure~\ref{fig:cryo}. A thin sample (1) is glued in the middle of an elastic cruciform polyimide substrate (2).  The geometry of the polyimide substrate is the same as in Ref. \cite{Gallofrantz2023}. Namely, the widths of the cross bars are 5 mm, the radii of the rounding are 2.5 mm. Each end of the cruciform substrate is clamped between brass blocks (4) and brass plates (3) by stainless still screws. In the center of each block (4) there is a threaded hole through which the threaded axis of the gearbox with 1:603 gear ratio rotated by a stepper motor (5) passes.
The motors with gearboxes are attached to the copper pot with a cold finger in the middle (6). This cold finger provides thermal connection between the sample (1) and temperature sensor (7) attached to the bottom of the pot (6). We tried different shapes of cold fingers and finally decided to use one with a flat $7\times 7$ mm$^2$ square on top since does not create additional tension in the middle of the substrate, like a round one. The pot is connected by a screw to a standard 24 pin panel (8). 
4 pins soldered to the top of the foiled fiberglass disk (7) are used to create an electrical connection to the sample (Fig. ~\ref{fig:both}(a)).
The copper pot with the motors sits on a copper foiled fiberglass disk (7) with conducting paths on its bottom. All 16 contacts (8 of motors, 4 of a sample and 4 of a temperature sensor) are soldered to the panel (10). The remaining eight contacts are used for thermal grounding. 
Each pair of step motors located opposite each other is connected in serial to one of two step motor controller to reduce the control current value. The maximum tensile force when placing the device into a cryosystem exceeds 30 N at a supply voltage of 30 V. The force can be at least doubled in the case of parallel connection of the step motors and even more enhanced by replacing the cryogenic 20-Ohm wiring by a highly conductive one. The outer diameter and the hight of the biaxial tensile cell are 58 mm and 30 mm respectively. A small triangle cutout in the foiled fiberglass disk (7) (Fig.~\ref{fig:both}(a)) provides access to a locking handle of the zero force connector.

\section{Cryogenic Setup for Biaxial Stretching Cell}
The cryogenic setup for biaxial tensile cell is based on a cryocooler with the Gifford-McMahon cooling circle (RDK-10-320). The advantage of such a cryocooler is the possibility to use the system in the geometry when the second stage cold head is directed upward. This provides very convenient loading the cell. 

The biaxial tensile cell is placed inside a brass chamber with internal diameter 60 mm (15). The chamber is hermetically closed with a lid (14) that is vacuum sealed by vacuum grease. The electrical connections inside the chamber are provided by zero force 24 pin connector (18) with 8 thermally grounded contacts for additional thermal connection. The chamber is connected by a tube (21) to a helium source and a pump system. 
We use helium with a few millibar pressure as a heat exchange gas inside the cell. The chamber is thermally connected to the second stage of the cryocooler (22) through the heater (21) and is surrounded by the radiation screen (13,17). The cup of the cryostat (10), cold screen cup (13) and chamber lid (14) can be easily removed to provide access to the cell.  A heater (21) between the second stage of the cryocooler (22) and the chamber (14,15) is used to control the temperature. The volume between the chamber (15) and the cryostat wall (16) is pumped out by turbomolecular pump. The minimum cooling time down to 10 K for our cryocooler is 3 hour. The minimum achievable temperature of the current setup is 9.5 K and can be significantly decreased with using modern cryocoolers with lower minimal temperature.

\begin{figure*}
\includegraphics[width=0.32\textwidth]{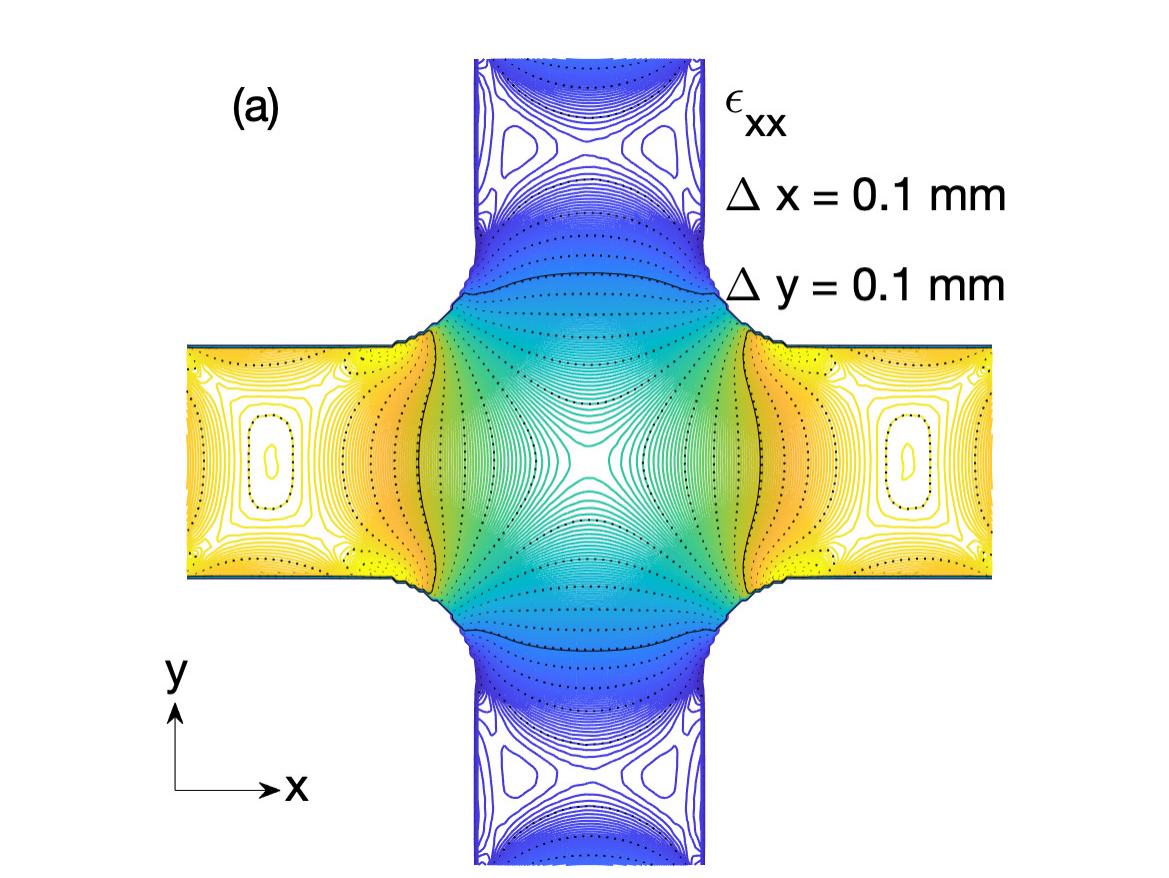}
\includegraphics[width=0.32\textwidth]{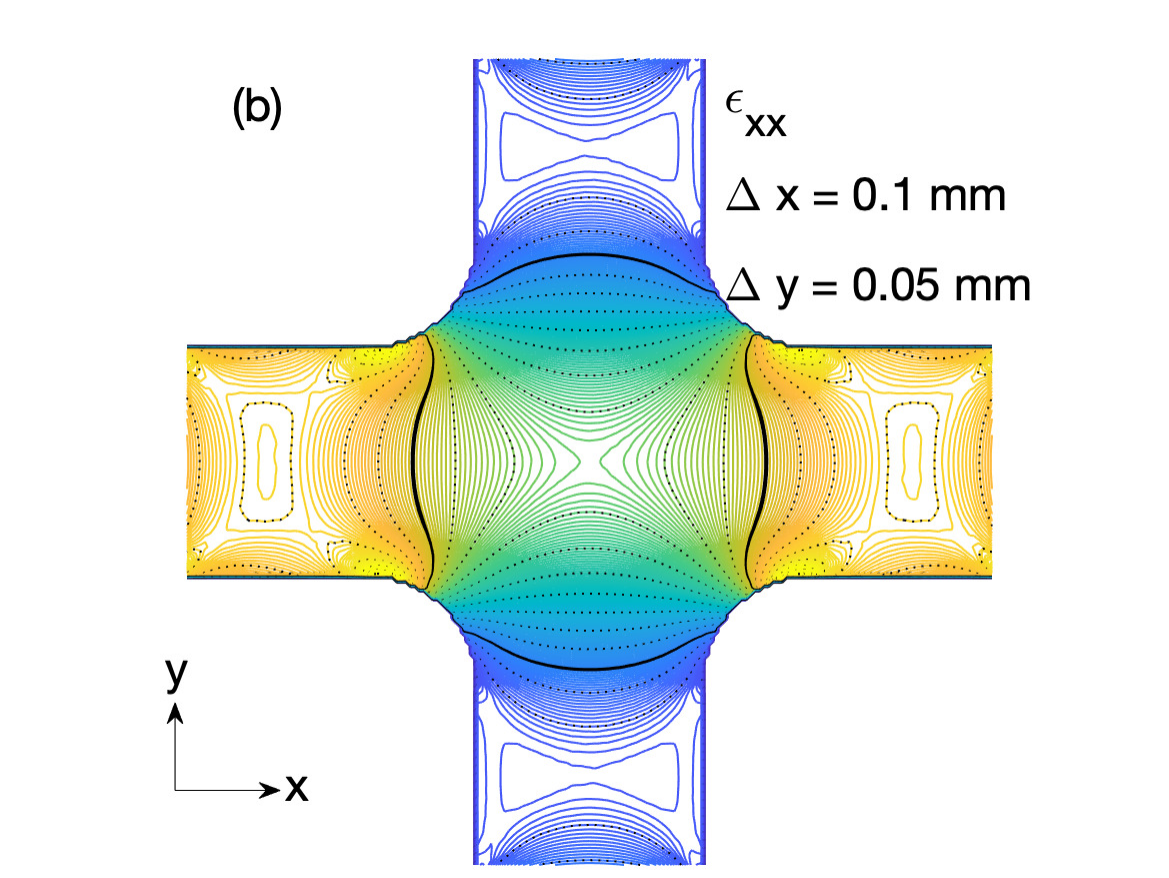}
\includegraphics[width=0.32\textwidth]{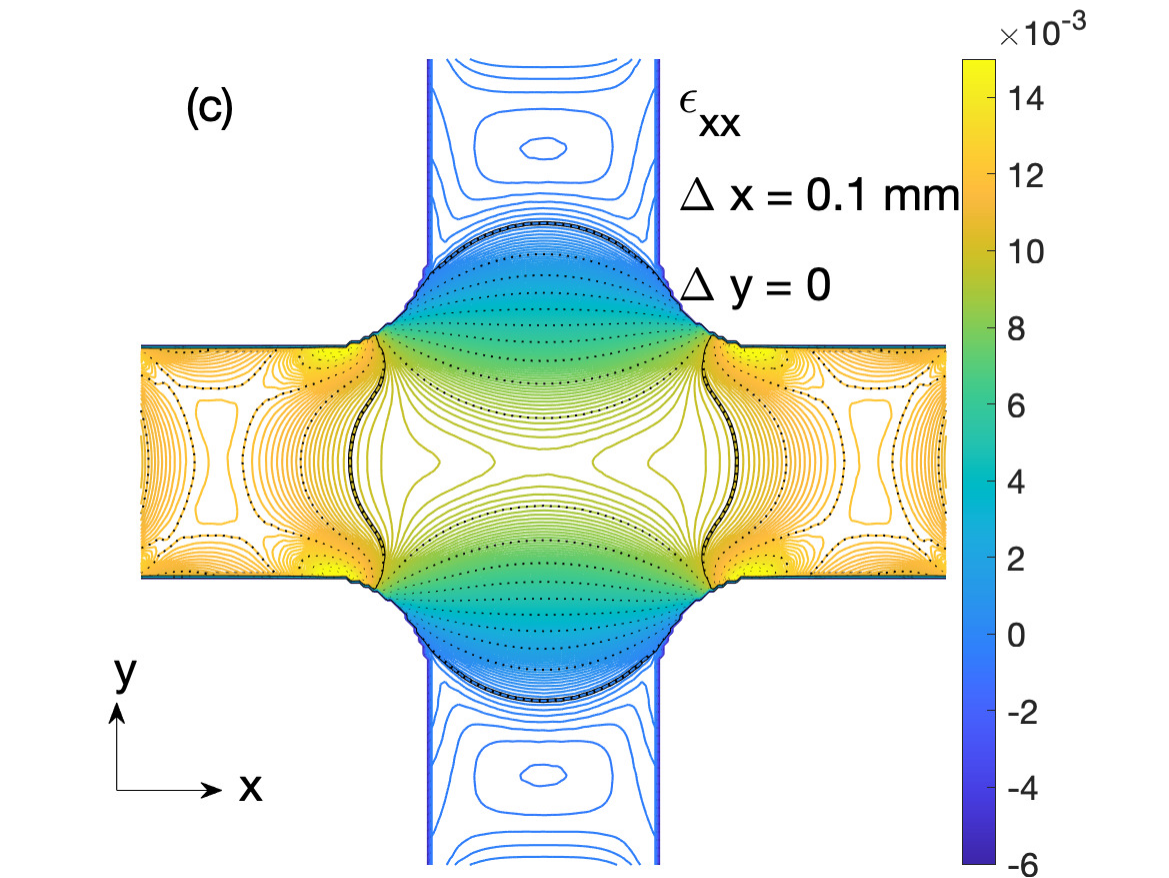}

\includegraphics[width=0.32\textwidth]{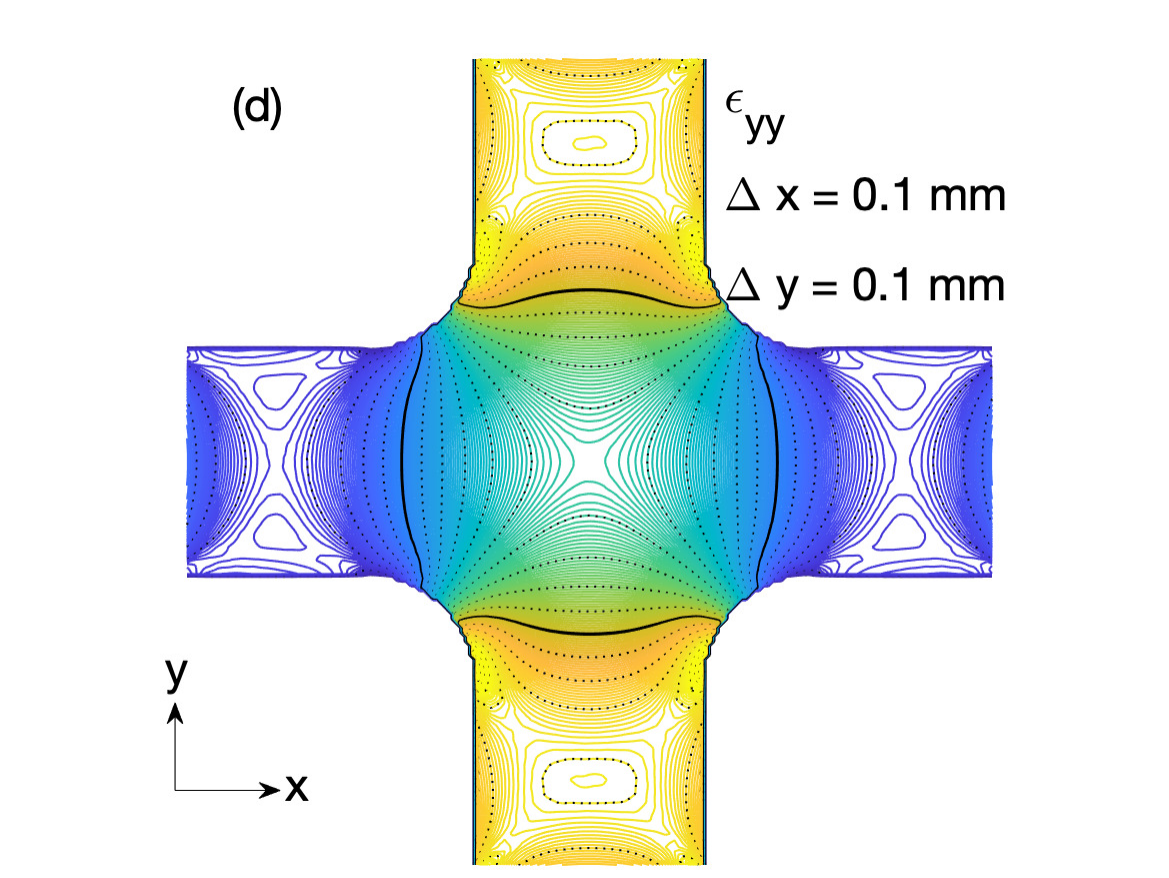}
\includegraphics[width=0.32\textwidth]{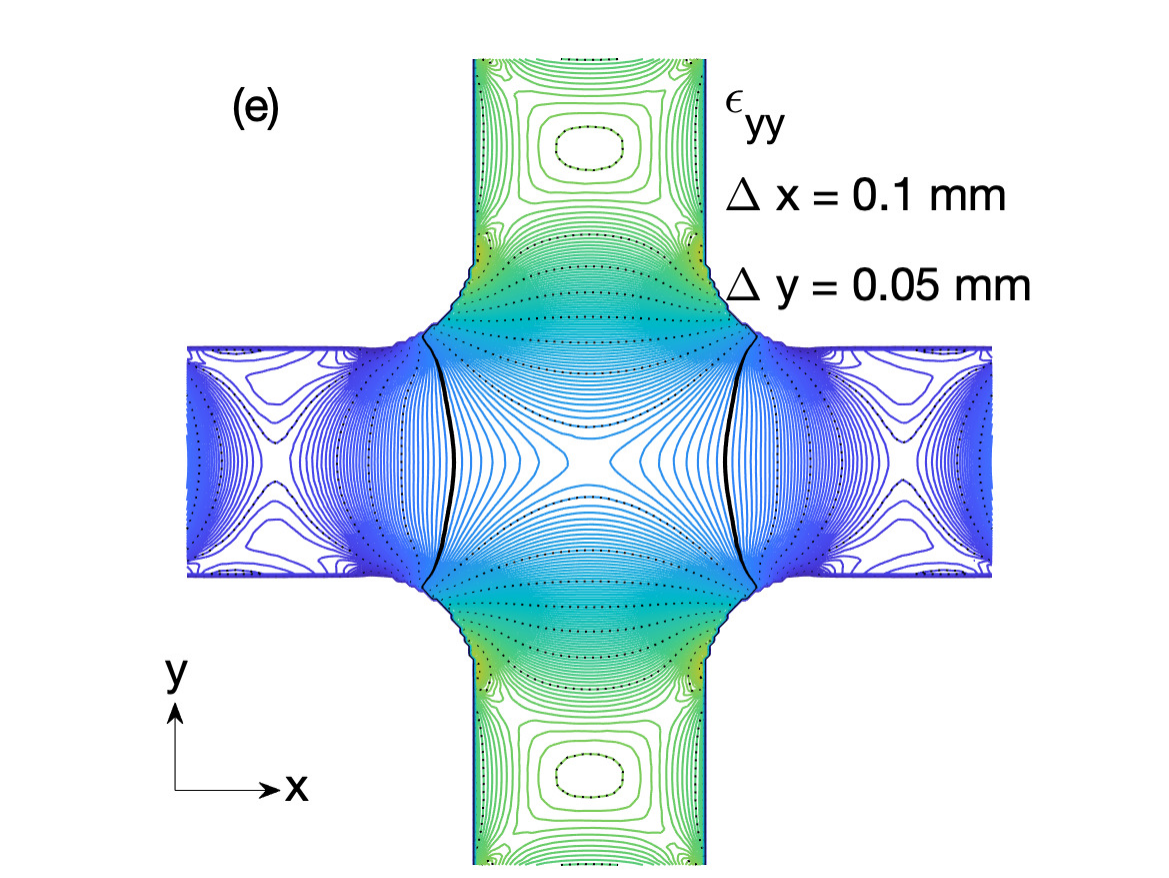}
\includegraphics[width=0.32\textwidth]{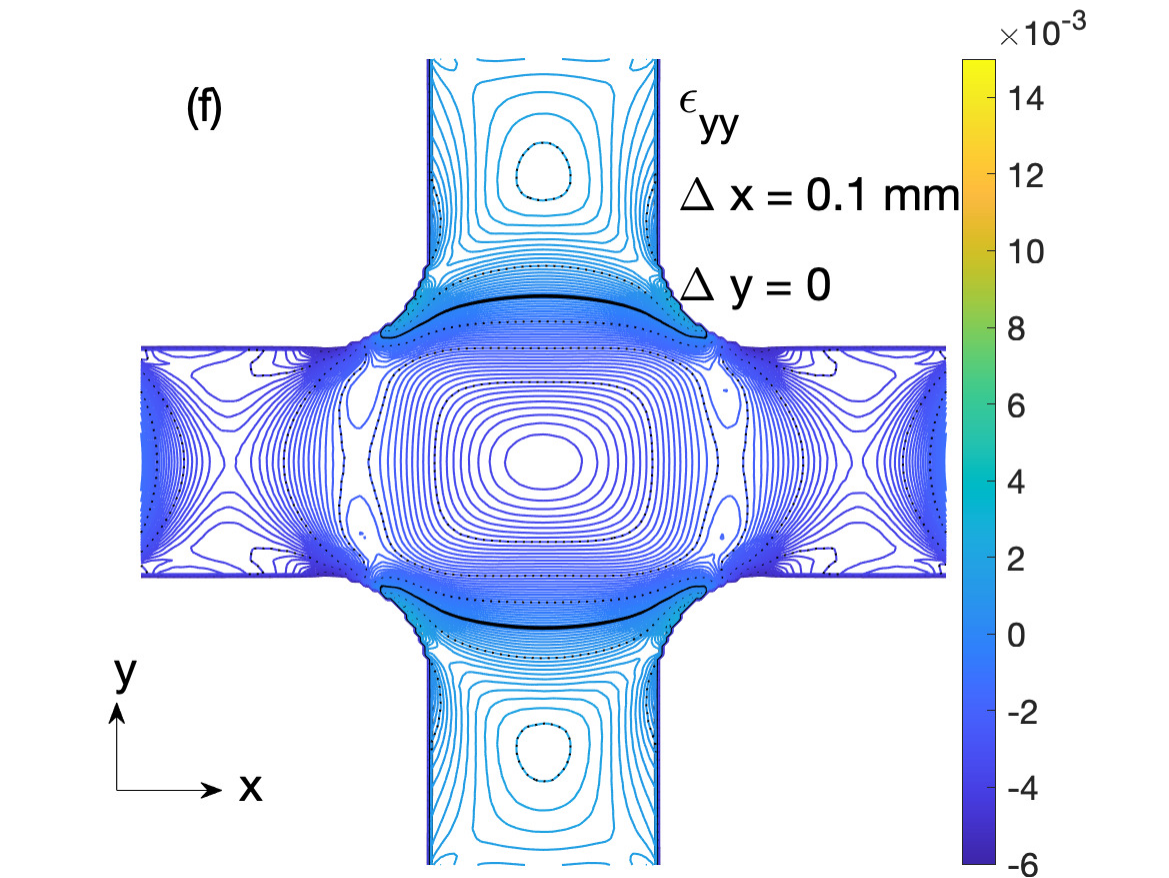}
\caption{Strain distributions in the polyimide cruciform substrate  with a thickness of 150 $\mu$m under various two-dimensional tensile loads. The strains of the cross ends are shown in each figure. All figures have the same color map shown in Fig. \ref{fig:stress}(c,f). Black lines show $\epsilon=0$ and 1\%, lines are drawn every 0.1\% \ and colored lines are drawn every 0.01\%.
\label{fig:stress}}
\end{figure*}

\section{Numerical Simulation of strain distribution}
The spatial distributions of two main components of the strain tensor, $\epsilon_{xx}$ and $\epsilon_{yy}$, were obtained by numerical methods using the standard finite element analysis of the substrate deformation. We use the finite-element simulation software based on COMSOL Multiphysics \cite{comsol}. The results described below are obtained under elastic approximation.  
The cruciform substrate shape and sizes used in the simulation ($18\times 18$~mm, width 5mm, rounding $R=2.5$~mm) corresponds to one used experimentally. We use fixed boundary conditions for cruciform ends and free one for other cruciform boundaries. The strain distribution was obtained for 2D geometry.

Following the official information provided by DuPont for polyimide (Kapton) \cite{Dupont}, we assume linear relation between the stress and strain and use the following parameters: Yang's modulus 2.97 GPa and Poisson's ratio 0.34. Note that the Poisson's ratio of polyimide increases from 0.34 at $\epsilon=1$\% \ to 0.48 \ at $\epsilon=1$5\%\cite{Bauer}.

\begin{figure}
\includegraphics[width=0.45\textwidth]{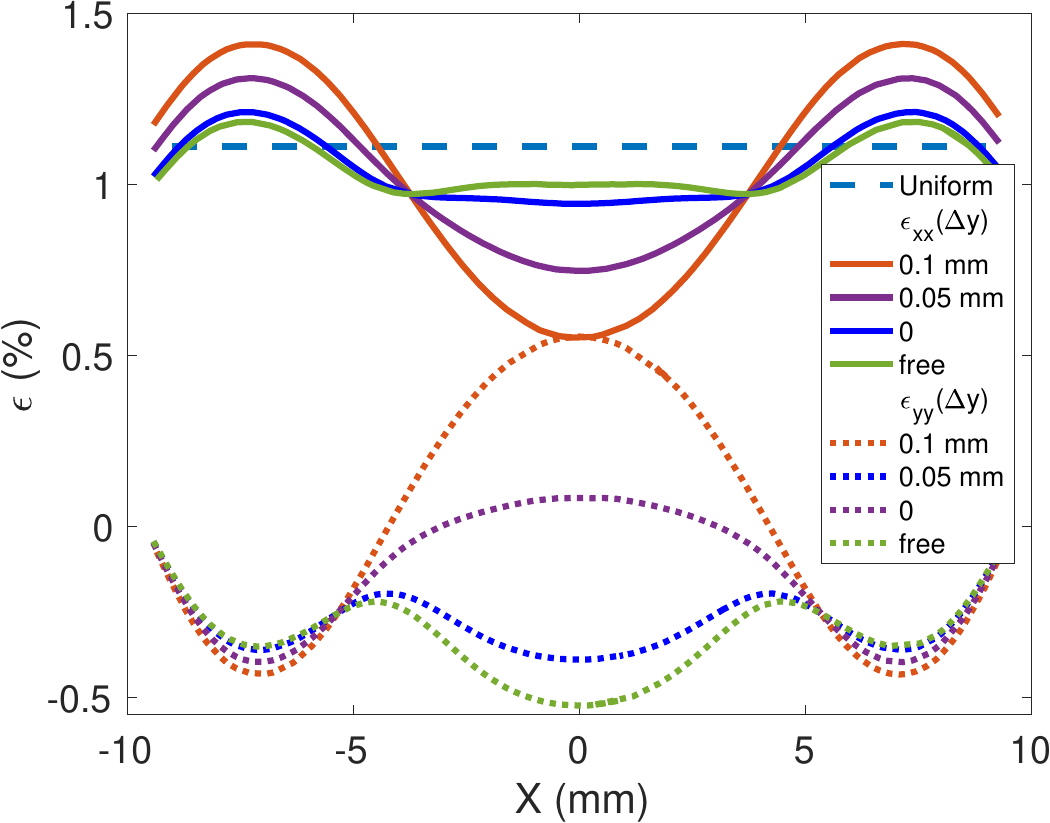}
\caption{Spatial distribution of two components of the strain tensor along the line passing the center of cruciform polyimide substrate at different values of tensile load along $x$ and $y$ axes.
\label{fig:strain}}
\end{figure}

Figure~\ref{fig:stress}(a-c) shows strain distributions $\epsilon_{xx}(x,y)$ when the ends of the horizontal crossbar are strained by $\Delta x =0.1$~mm away from the cross center in the $x$ and $y$ directions by $\Delta y = 0.1$~mm, 0.05 mm and 0 along $y$ direction. Figure~\ref{fig:stress}(d-f) shows similar distributions for $\epsilon_{yy}(x,y)$. Figures~\ref{fig:stress}(a,d) differ by $90^o$ rotation because $\Delta x=\Delta y$. Increase of $\Delta y$ affects $\epsilon_{xx}$ because of nonzero Poisson's ratio. 
The mutual influence of two orthogonal deformation components is better visible in the figure~\ref{fig:strain} which shows strain distributions along the horizontal line passing the center of the cross obtained with different values of vertical elongation $\Delta y$. We see that $\epsilon_{xx}$ in the center of the cruciform substrate ($x=0$,$y=0$) has a local minimum and significantly depends on elongation in the vertical direction. Namely, increase of $\Delta y$ from 0 to $\Delta y = \Delta x= 0.1$~mm results in 40\% decrease of $\epsilon_{xx}$.  If we want to keep purely uniaxial strain along either $x$ or $y$ axes, that corresponds to $\epsilon_{yy}=0$ or $\epsilon_{xx}$=0 respectively, application of 41\% displacements along the orthogonal direction is required.

\begin{figure}
\includegraphics[width=0.42\textwidth]{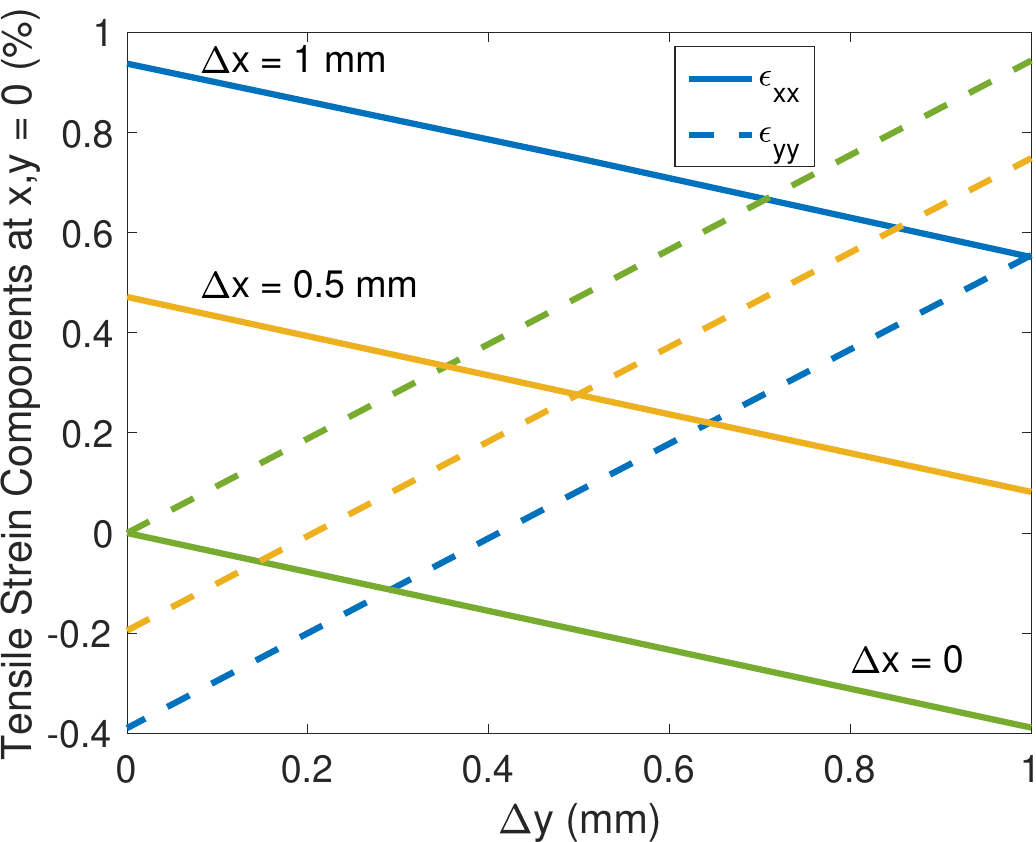}
\caption{Components of the strain tensor in the  center of the cruciform polyimide substrate with 3 different tensile loads along $\Delta x=1$~mm as a function of tensile load along $y$ axis.
\label{fig:yload}}
\end{figure}

Figure~\ref{fig:yload} gives a more detailed view of this dependence. Namely, it shows the strain components $\epsilon_{xx}$ and $\epsilon_{yy}$ in the  center of the polyimide cruciform substrate with tensile load $\Delta x=0.1$~mm as a function of tensile load along $y$ axis. This dependence is nonmonotonous indeed and reachs its minimum near $\Delta y = 0.6$~mm. Such a complex behavior is a consequence of nonzero value of the Poisson's ratio. That means, that in a real bidirectional strain study, it is preferable to control the strain components of a sample by independent methods.

\section{Bidirectional measurement technique}
\begin{figure}[h]
\includegraphics[width=0.48\textwidth]{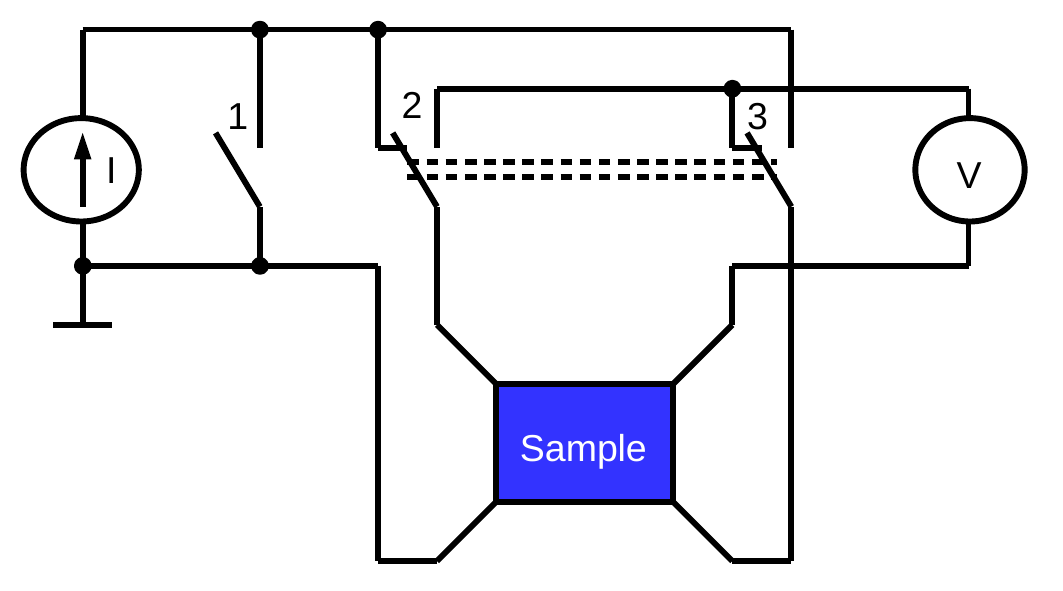}%
\caption{Scheme of sample resistance measurements. Relays 2 and 3 are used to
switch the direction of the current. Relay 1 provides a short circuit for the current
source when reconnecting the sample. \label{fig:scheem}}%
\end{figure}

The possibility to deform a crystal in two directions makes it desirable to perform simultaneous measurements of at least two components of the conductivity tensor. This problem can be solved by using a computer controlled relay board as it is shown in Figure~\ref{fig:scheem}. Resistance along each direction is measured sequentially by reconnecting sample contact between two geometries when the current flows along the vertical direction (Fig.~\ref{fig:scheem}, relays 2 and 3 are both off as is shown) and the horizontal one (relays 2 and 3 are both on). Relay 1 is used to create a short circuit to exclude breaking of the current source circuit on reconnection of the measurement direction. This allows us to exclude voltage jumps on the sample. In our case, the resistance measurements are carried out by either using the EGG7265 lock-in amplifier, which is also used to control an Arduino relay board using its digital output, or SRS 830 lock-in amplifier using its DACs outputs to control the relay board. The data obtained by these methods can be used for calculation of the diagonal components of conductivity tensor by Montgomery method \cite{Montgomery_2003}. There is also another algorithm for using this circuit, in which only relays 2 and 3 are used. In this case, the switching sequence of relays 2 and 3 should ensure a constant short circuit of the current source to the sample. Namely, for the circuit shown in the Fig.~\ref{fig:scheem}, to measure conductivity in the horizontal direction, you first need to switch relay 3, and only then relay 2. To switch back to measure conductivity in the vertical direction, you must first switch relay 2, and only then relay 3.

\section{Test of the biaxial tensile cell}
\subsection{Strain components measurements}
Experimental demonstration the effect of deformation along different directions is done by using quasi-one dimensional metal TaSe$_3$ as a strain sensor. The resistivity $\rho$ of this material is very sensitive to unidirectional strain $\epsilon$ \cite{Tritt} and depends exponentially on the strain, $\rho=\rho_0e^{a\epsilon}$, where $a\approx 10^{-2}$  \ \cite{Lukmanova} and $\rho_0(300{\rm \ K}) \approx 10^{-3}$~$\Omega\cdot$cm \cite{TaSe3rho}. In the present study, a thin whisker of TaSe$_3$ was glued in the middle of the cruciform polyimide substrate along $x$ direction by a silver paint or attached to the surface by cold soldering of thin indium wires.  Polyimide substrate thickness was 150~$\mu$m. In such a geometry it is sensitive to $x$-component of the strain only. Its resistance was measured by four-probe technique at room temperature by a lock-in amplifier as a function of symmetrical tensile load  along $x$ direction only ($\Delta y=0$). This geometry corresponds to Fig.~\ref{fig:stress}(c,f). The measurements were carried out at room temperature in the current control mode at a modulation frequency of 7 Hz.

\begin{figure}
\includegraphics[width=0.44\textwidth]{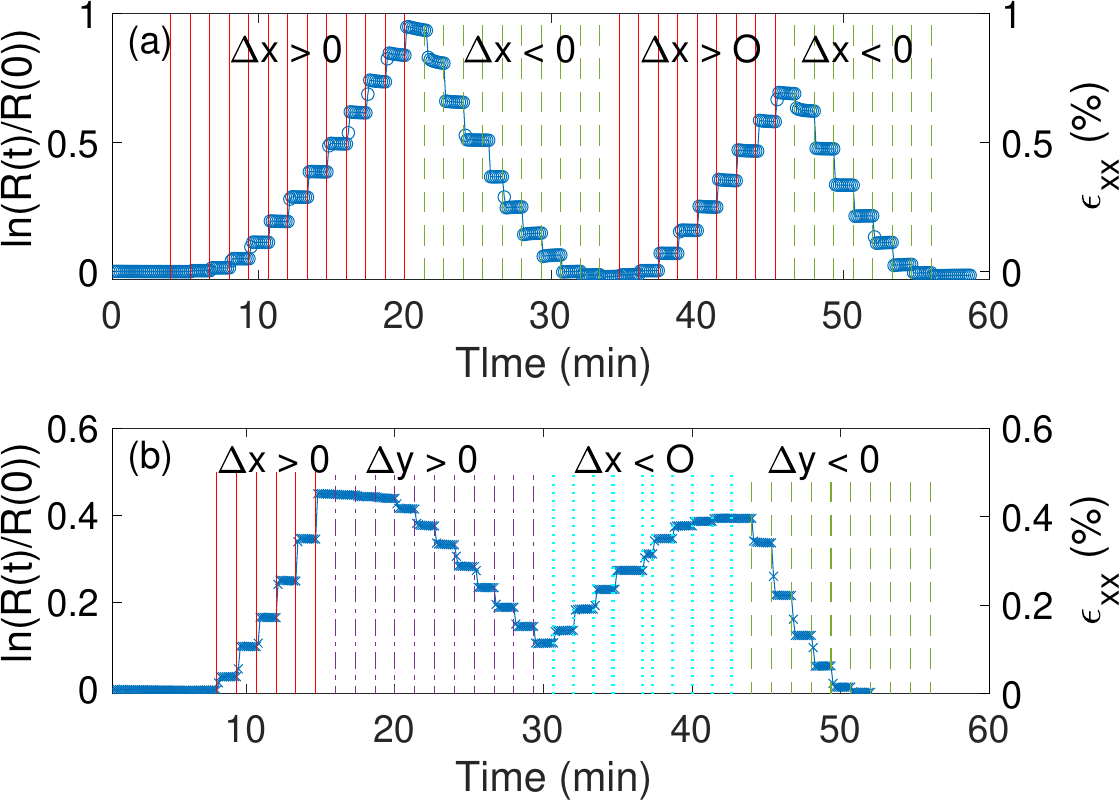}%
\caption{(a) (a) Room temperature sample resistance evolution due to back and force stepwise changes of the cruciform substrate length along the $x$ direction. The red vertical solid lines correspond to an increase in the cross length, and the green dashed lines correspond to a decrease in the length. (b) The same when the cross length is changing along both $x$ and $y$ directions. The red vertical solid lines correspond to an increase in the cross length, and the green dashed lines correspond to a decrease in the length. The magenta dashed-dotted and cyan dotted lines correspond to increases and decreases in the y direction, respectively. \label{fig:results}}%
\end{figure}

Figure~\ref{fig:results}(a) shows time evolution of the resistance resulted from tensile steps of the substrate along $x$ direction only. Red solid vertical lined corresponds to steps in the strain increase direction, whereas green dashed lined marks the steps in the opposite direction. The right axis shows the strain estimation following Ref. \cite{Lukmanova}. In the initial stage, the substrate is relaxed in both directions. So the first few steps along $x$ axis gives no response (Fig.~\ref{fig:results}(a)). Then the respond appears.  Further progress in the initial direction leads to smooth increase of the step heights. Reverse of the step direction results in change of the step directions. Small relaxation of the strain is seen when the sensor resistance doubles  ($\epsilon_{xx}\approx 0.7$). The reverse of strain change directions at $t=20$~min gived somewhat bigger steps with much weaker relaxation. The second run with smaller maximum strain value ($t>35$~min) demonstrate similar behavior except for the absence of noticeable relaxation.

Fig.~\ref{fig:results}(b) illustrates evolution of the resistance resulted from tensile steps of the substrate along both $x$ and $y$ directions. The first 6 steps (red solid lines) are used to produce initial strain along $x$ direction. Then similar steps are done in $y$ direction (violet dash-dotted lines). when the response appears, it is directed in the opposite way and reduces $\epsilon_{xx}$. Reversing direction of the steps (cyan dotted lines) increases $\epsilon_{xx}$. The mean step height ratio gives $(d\epsilon_{xx}/dx)/(d\epsilon_{yy}/dy)=-0.4$-$-0.5$, in agreements with numerical simulations results shown in Figure~\ref{fig:yload}. Such a behavior provides additional possibilities in experimental study of biaxial strain effects.

\subsection{Testing with 2D sample}

The main purpose of the developed device is to study the effect of biaxial deformation on temperature dependent transport of layered materials. For demonstration purposes, 
we have prepared a sample of a layered compound 2H-NbS$_2$ of approximately rectangular shape with $L_1\times L_2=0.82 \times 0.38 $ mm$^2$ lateral sizes and a thickness of about 10 microns. Polyimide substrate thickness was 75~$\mu$m. The crystal structure of 2H-NbS$_2$ consists of hexagonal S-Nb-S sandwiches ($a-b$ plain) coupled through weak van der Waals interaction along $c$ direction. The sample thickness was estimated from a sample resistance with room-temperature resistivity $\rho_{a,b}=1.5\cdot 10^{-4}$~$\Omega\cdot$cm \cite{NbS2}). As the sample and substrate thicknesses are comparable, we do not expect the sample deformation to be equal to the substrate deformation, but we expect that they will be proportional to each other. The later is enough for test purposes. The sample was oriented to have edges along the cross bars and glued to the middle of the substrate.The electrical contacts to the sample were deposited on the top surface of the sample corners by physical vapor deposition of indium. Wires to the deposited indium contacts were attached by a silver paint. 

Large sizes of the tensile cell and relatively long thermal contact between the sample and temperature sensor makes actual study of accuracy of temperature measurements. For this purpose we have undertaken studies of $R_i(T)|_{i=1,2}$ with different temperature sweep direction. In our tensile cell, if the measurements are performed with a constant sweep rate $\leq 1$~K/min and the heat exchange gas pressure 3-5 mbar, the thermal hysteresis loop width $\Delta T$ does not exceed 0.7\%$T$ over entire temperature range. 

\begin{figure}
\includegraphics[width=0.44\textwidth]{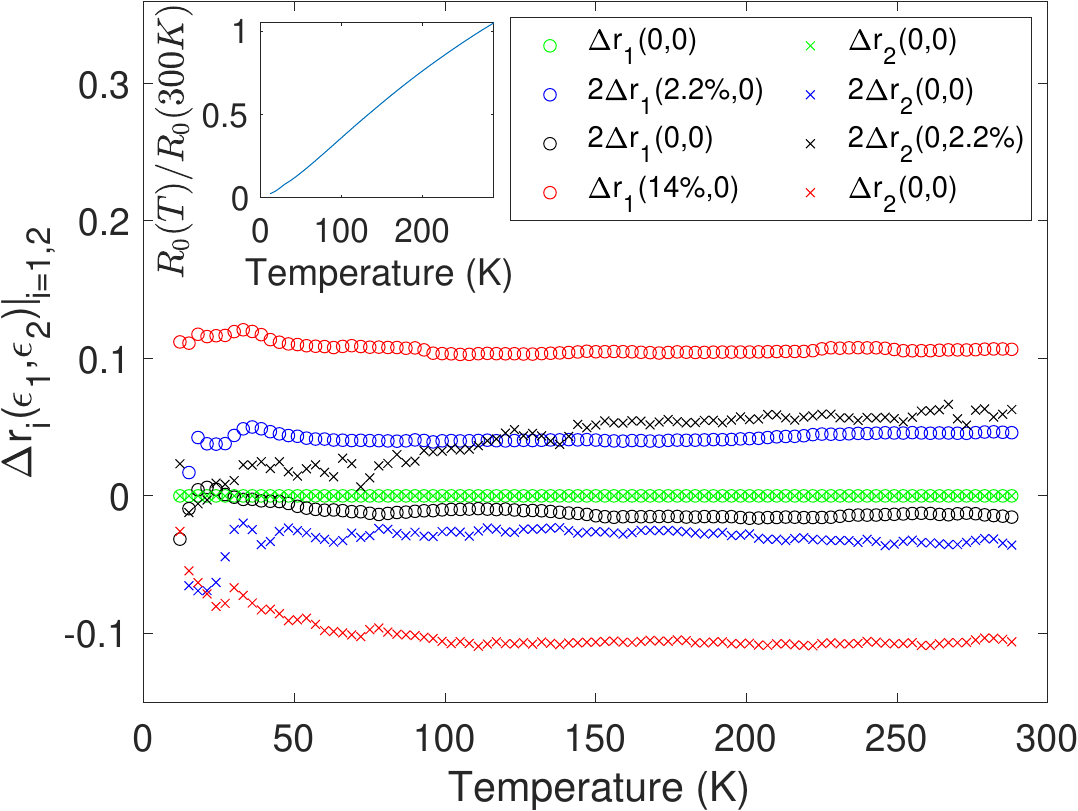}%
\caption{Effect of strain on temperature dependences of normalized sample resistances along two directions. Inset shows temperature variation of the sheet resistivity of undeformed crystal. \label{fig:2DRT}}%
\end{figure}

Resistances along two orthogonal directions of strain application $R_1(T)$ and $R_2(T)$ were measured using the switching circuit described above. The measurements were carried out in the current control mode at a modulation frequency of 7 Hz. Temperature dependence of the sheet resistance $R_0(T)$ in the initial state  has the usual form (inset (a) in Fig.~\ref{fig:2DRT}) and corresponds to the results published earlier \cite{NbS2}. When the sample is deformed by strain application to the polyimide substrate, the resistivity increases along strain direction and decreases along perpendicular one, whatever the strain direction is (Fig.~\ref{fig:2DRT}). Temperature dependence of $\Delta r_i =R_i(\epsilon,T)/R_i(0,T)|_{i=1,2}-1$ (here $\epsilon$ is relative elongation of the cross which is controlled by counting the steps of the step motor) remains nearly constant at $T\gtrsim 100$~K. At lower temperature, there is a noticeable temperature dependence for $\delta r_2$ measured along shorter sample length. We were able to obtain $\Delta r_i(T)|_{i=1,2}\approx 10$\% without noticeable damage of the sample.  Our attempt to produce larger deformation was limited by plastic deformation of the sample.

The observed behavior corresponds to general expectations. For instance, in the tight binding model, this type of behavior corresponds to expected suppression of overlap integrals along strain direction and their increase in the transverse one. The resistivity anisotropy of 2H-NbS$_2$ for the transverse $c$ direction is very high and varies from $A\equiv \rho_c/ \rho_{a,b}=2\cdot 10^3$ at room temperature to $\rho_c/ \rho_{a,b}=7\cdot 10^3$ at $T\approx 10$~K \cite{NbS2}. That means that our sample cannot be considered as a 2D conducting object. Therefore, since transverse conductivity is also affected by two-dimensional strain due to non-zero Poisson's ratio, its contribution to the measured resistance also depends on strain and temperature and increases with increasing $A$, i.e. with decreasing temperature. This effect should be more pronounced for closely spaced contacts, which is consistent with our observations. Further study is required to clarify the origin of the observed low-temperature behavior. 
 
\section{Concluding remarks}
Knowledge of the geometries of an elastic substrate and the tensile system allows one to estimate the expected deformation of the elastic substrate. However, this estimate is very rough since it does not take into account the unknown thickness and homogeneity of the glue with which the sample is glued, the influence of contacts, or the sample itself. Therefore, the most correct method is the direct measurement of crystal lattice parameters using x-ray analysis, as performed in Ref..\cite{Gallofrantz2023}. The tensile cell described above can be easily used in X-ray measurement and Raman study. 

Another problem is the thermal expansion of setup components, which may additionally affect sample deformation during temperature sweep. This effect works differently for the tensile setup described in Ref.\cite{Gallofrantz2023} and here. In Ref.\cite{Gallofrantz2023}, the stain force applied to
the substrate is controlled over the entire temperature range. So the temperature dependence of Yang’s modulus, Poisson’s ratio, and thermal expansion of a substrate affect the sample deformation value. We would like to note that many plastics shrink a lot in cryogenic environments. In the present setup, we control substrate deformation only with respect to the copper cell body. Thus, deformation is affected mostly by the temperature expansion of the copper. If we assume that the thermal expansion of a studied material is not anomalous and close to typical values of many other constructive materials, including copper, then we could expect that the sample strain stays approximately the same with temperature variation. Fortunately, the thermal expansion coefficient of polyimide is very close to that of usual materials, at least at room temperature ($2\cdot 10^{-5}$~1/K\cite{Dupont}). The results obtained demonstrate that this is indeed a reasonably good practical approximation. However, in both cases of applied force and substrate stain control, independent control of the sample strain over the entire temperature range is desirable.

{\bf In conclusion}, we describe here the compact biaxial tensile cell with independent strain control along two orthogonal directions. The cell is designed for electrical measurements at cryogenic temperatures but can also be used for a wide range of studies using other methods..

\section{Data availability}
The data that support the finding of this study are available  from the corresponding author on reasonable request.

\begin{acknowledgments}
The financial support from the RScF project \# 22-42-09018 is acknowledged.
\end{acknowledgments} 

\bibliography{strain_device}

\end{document}